\documentclass[aps, pra, amsmath, amssymb, reprint, superscriptaddress, longbibliography]{revtex4-1}

\pdfoutput=1

\usepackage{graphicx}
\usepackage{dcolumn}% Align table columns on decimal point
\usepackage{bm}% bold math
\usepackage{upgreek}
\usepackage{xspace}
\usepackage[caption=false]{subfig}

\newcommand{\ARD}{\ensuremath{A_\mathrm{RD}}\xspace}
\newcommand{\Ab}{\ensuremath{A_\mathrm{b}}\xspace}
\newcommand{\ceff}{\ensuremath{c_\mathrm{eff}}\xspace}
\newcommand{\dec}{\ensuremath{d_\mathrm{ec}}\xspace}
\newcommand{\Ea}{\ensuremath{E_\mathrm{a}}\xspace}
\newcommand{\Eg}{\ensuremath{\Delta}\xspace}
\newcommand{\Enull}{\ensuremath{E_0}\xspace}
\newcommand{\etaC}{\ensuremath{\eta_\mathrm{C}}\xspace}
\newcommand{\etacc}{\ensuremath{\eta_\mathrm{cc}}\xspace}
\newcommand{\etal}{\ensuremath{\eta_\mathrm{l}}\xspace}
\newcommand{\etamax}{\ensuremath{\eta_\mathrm{max}}\xspace}
\newcommand{\etas}{\ensuremath{\eta_\mathrm{s}}\xspace}
\newcommand{\Evac}{\ensuremath{E_\mathrm{vac}}\xspace}
\newcommand{\fFD}{\ensuremath{f_\mathrm{FD}}\xspace}
\newcommand{\Ibemax}{\ensuremath{I_\mathrm{be}^\mathrm{max}}\xspace}
\newcommand{\Iecmax}{\ensuremath{I_\mathrm{ec}^\mathrm{max}}\xspace}
\newcommand{\Iec}{\ensuremath{I_\mathrm{ec}^\mathrm{}}\xspace}
\newcommand{\Iemax}{\ensuremath{I_\mathrm{e}^\mathrm{max}}\xspace}
\newcommand{\Ig}{\ensuremath{I_\mathrm{g}}\xspace}
\newcommand{\Inull}{\ensuremath{I_\mathrm{0}}\xspace}
\newcommand{\Iout}{\ensuremath{I_\mathrm{out}}\xspace}
\newcommand{\IRD}{\ensuremath{I_\mathrm{RD}}\xspace}
\newcommand{\Jbemax}{\ensuremath{J_\mathrm{be}^\mathrm{max}}\xspace}
\newcommand{\JbeEnullmax}{\ensuremath{J_\mathrm{be,E0}^\mathrm{max}}\xspace}
\newcommand{\JEnull}{\ensuremath{J_\mathrm{E0}}\xspace}
\newcommand{\Jecmax}{\ensuremath{J_\mathrm{ec}^\mathrm{max}}\xspace}
\newcommand{\JeEnullmax}{\ensuremath{J_\mathrm{e,E0}^\mathrm{max}}\xspace}
\newcommand{\Jemax}{\ensuremath{J_\mathrm{e}^\mathrm{max}}\xspace}
\newcommand{\Jepete}{\ensuremath{J_\mathrm{e}^\mathrm{PETE}}\xspace}
\newcommand{\JRD}{\ensuremath{J_\mathrm{RD}}\xspace}
\newcommand{\kB}{\ensuremath{k_\mathrm{B}}\xspace}
\newcommand{\me}{\ensuremath{m_\mathrm{e}}\xspace}
\newcommand{\muc}{\ensuremath{\mu_\mathrm{c}}\xspace}
\newcommand{\mue}{\ensuremath{\mu_\mathrm{e}}\xspace}
\newcommand{\mycelsius}{^\circ\hspace{-0.08em}C}
\newcommand{\mytilde}{\raise.17ex\hbox{$\scriptstyle\mathtt{\sim}$}\,}
\newcommand{\myunit}[2]{\ensuremath{#1\,\mathrm{#2}}\xspace}
\newcommand{\Pb}{\ensuremath{P_\mathrm{b}}\xspace}
\newcommand{\Pcond}{\ensuremath{P_\mathrm{cond}}\xspace}
\newcommand{\Pcondc}{\ensuremath{P_\mathrm{condc}}\xspace}
\newcommand{\Pel}{\ensuremath{P_\mathrm{el}}\xspace}
\newcommand{\Pelc}{\ensuremath{P_\mathrm{elc}}\xspace}
\newcommand{\Pg}{\ensuremath{P_\mathrm{g}}\xspace}
\newcommand{\phie}{\ensuremath{\phi_\mathrm{e}}\xspace}
\newcommand{\phig}{\ensuremath{\varphi_\mathrm{g}}\xspace}
\newcommand{\phic}{\ensuremath{\phi_\mathrm{c}}\xspace}
\newcommand{\Pin}{\ensuremath{P_\mathrm{in}}\xspace}
\newcommand{\Pout}{\ensuremath{P_\mathrm{out}}\xspace}
\newcommand{\Prad}{\ensuremath{P_\mathrm{rad}}\xspace}
\newcommand{\Pradc}{\ensuremath{P_\mathrm{radc}}\xspace}
\newcommand{\Prej}{\ensuremath{P_\mathrm{rej}}\xspace}
\newcommand{\Psimax}{\ensuremath{\Psi_\mathrm{max}}\xspace}
\newcommand{\Rl}{\ensuremath{R_\mathrm{l}}\xspace}
\newcommand{\Rle}{\ensuremath{R_\mathrm{le}}\xspace}
\newcommand{\Rlc}{\ensuremath{R_\mathrm{lc}}\xspace}
\newcommand{\Tc}{\ensuremath{T_\mathrm{c}}\xspace}
\newcommand{\Te}{\ensuremath{T_\mathrm{e}}\xspace}
\newcommand{\Tin}{\ensuremath{T_\mathrm{in}}\xspace}
\newcommand{\Ts}{\ensuremath{T_\mathrm{s}}\xspace}
\newcommand{\Tout}{\ensuremath{T_\mathrm{out}}\xspace}
\newcommand{\Tnull}{\ensuremath{T_\mathrm{0}}\xspace}
\newcommand{\upd}{\ensuremath{\mathrm{d}}\xspace}
\newcommand{\Vout}{\ensuremath{V_\mathrm{out}}\xspace}
\newcommand{\Vlead}{\ensuremath{V_\mathrm{lead}}\xspace}
\newcommand{\Vge}{\ensuremath{V_\mathrm{ge}}\xspace}

\begin{document}

\title[Highly-Efficient Thermoelectronic Conversion of Solar Energy and Heat into Electric Power]{Highly-Efficient Thermoelectronic Conversion of Solar Energy and Heat into Electric Power}

\author{S. Meir}
\affiliation{Center for Electronic Correlations and Magnetism, Experimental Physics VI, Augsburg University, 86135 Augsburg, Germany}
\author{C. Stephanos}
\affiliation{Center for Electronic Correlations and Magnetism, Experimental Physics VI, Augsburg University, 86135 Augsburg, Germany}
\affiliation{Max Planck Institute for Solid State Research, 70659 Stuttgart, Germany}
\author{T.H. Geballe}
\affiliation{Department of Applied Physics and Laboratory for Advanced Materials, Stanford University, Stanford, CA 94305-4045, USA}
\author{J. Mannhart}
\email[author to whom correspondence should be addressed, ]{j.mannhart@fkf.mpg.de}
\affiliation{Max Planck Institute for Solid State Research, 70659 Stuttgart, Germany}

\date{2013-01-15}

\begin{abstract}
Electric power may, in principle, be generated in a highly efficient manner from heat created by focused solar irradiation, chemical combustion, or nuclear decay by means of thermionic energy conversion. As the conversion efficiency of the thermionic process tends to be degraded by electron space charges, the efficiencies of thermionic generators have amounted to only a fraction of those fundamentally possible. We show that this space-charge problem can be resolved by shaping the electric potential distribution of the converter such that the static electron space-charge clouds are transformed into an output current. Although the technical development of practical generators will require further substantial efforts, we conclude that a highly efficient transformation of heat to electric power may well be achieved.
\end{abstract}

\maketitle

\section{Introduction}

Electric power can be generated in a highly efficient manner via thermionic energy conversion from heat created by focused solar irradiation or combustion of fossil fuels~\cite{schlichter_spontane_1915,ingold_calculation_1961,hatsopoulos_thermionic_1973,schwede_photon-enhanced_2010}. Generators based on the thermionic process could, if implemented, considerably enhance the efficiency of focused solar energy conversion or of coal combustion power plants~\cite{fitzpatrick_updated_1997}, yielding a corresponding reduction of CO$_2$ emissions. In thermionic energy conversion a vacuum is applied as the active material between the electrodes, rather than the solid conductors that give rise to the thermoelectric effect~\cite{ioffe_semiconductor_1957}. Thereby, the parasitic heat conduction from the hot to the cold electrode is radically decreased.

Thermionic generators can operate with input temperatures \Tin that are sufficiently high to match the temperatures at which concentrating-solar power plants or fossil-fuel power stations generate heat. In principle, electric power may therefore be generated from these energy sources with outstanding efficiency because the maximum possible efficiency -- the Carnot efficiency $\etaC=1-\frac{\Tout}{\Tin}$ -- increases with \Tin, where \Tout is the generator‘s output temperature. In contrast, a significant amount of energy is wasted today in the conversion of heat to electricity. Coal, from which \myunit{40}{\%} of the world‘s electricity is currently generated~\cite{_key_2012}, is burned in power stations at $\mytilde \myunit{1500}{\mycelsius}$, whereas, due to technical limitations, the steam turbines driven by this heat are operated below $\mytilde \myunit{700}{\mycelsius}$, to give but one example.

However, thermionic generators have never been deployed to harvest solar energy or to convert combustion heat into electricity in power stations~\cite{engdahl_thermionics_1970} or cars~\cite{moyzhes_thermionic_2005}, although the conversion process is straightforward and appears to be achievable: electrons are evaporated from a heated emitter electrode into vacuum, then the electrons drift to the surface of a cooler collector electrode, where they condense~\cite{hatsopoulos_thermionic_1973,ioffe_semiconductor_1957}. If used for solar energy harvesting, the quantum nature of light can be exploited for great efficiency gains by using photon-enhanced thermionic emission (PETE)~\cite{schwede_photon-enhanced_2010}. PETE employs the photoeffect to enhance electron emission by lifting the electron energy in a semiconducting emitter across the bandgap $\Delta$ into the conduction band, from where the electrons are thermally
emitted. As a result of the electron flow, the electrochemical potentials of the emitter and collector differ by a voltage \Vout, and an output current $\Iout=\Vout/\Rl$ can be sourced through a load resistor \Rl. Turning this elegant operation principle into commercial devices has not yet been possible, however, because space-charge clouds suppress the emission current for emitter-collector distances of $\dec > 3$--$\myunit{5}{\upmu m}$~\cite{hatsopoulos_measured_1958,ioffe_semiconductor_1957,moss_thermionic_1957}. Practical fabrication of emitter-collector assemblies that operate with the required close tolerances at a temperature difference $\Te-\Tc$ of many hundred Kelvin was found to be highly challenging~\cite{national_research_council._committee_on_thermionic_research_and_technology._thermionics_2001}. In addition, for $\dec < \myunit{1}{\upmu m}$, near-field infrared thermal losses between emitter and collector become large~\cite{lee_optimal_2012}. For large \dec, it has only been possible to suppress the space charges by neutralizing them, which was done by inserting Cs$^+$ ions into the space-charge cloud~\cite{rasor_thermionic_1991,rasor_emission_1963}, a method used in two \myunit{5}{kW} nuclear-powered thermionic generators aboard experimental Soviet
satellites~\cite{national_research_council._committee_on_thermionic_research_and_technology._thermionics_2001,ponomarev-stepnoi_russian_2000}. With that approach, compensating the space charge by ion injection causes a $\mytilde\myunit{50}{\%}$ loss of output power \Pout~\cite{moyzhes_thermionic_2005}. Novel schemes to suppress the space charges by optimizing the generation of Cs$^+$~\cite{moyzhes_thermionic_2005} have yet to be demonstrated. Since the 1950s, when the space-charge problem was first approached~\cite{baksht_thermionic_1978,hatsopoulos_thermionic_1973,ioffe_semiconductor_1957}, it has remained the main obstacle to achieving efficient thermionic generators~\cite{hatsopoulos_thermionic_1973,moyzhes_thermionic_2005}.

\section{Resolving the space-charge problem with electric and magnetic fields }

\begin{figure}
  \subfloat[]{\hspace*{-.04\columnwidth}\includegraphics[width=1.05\columnwidth]{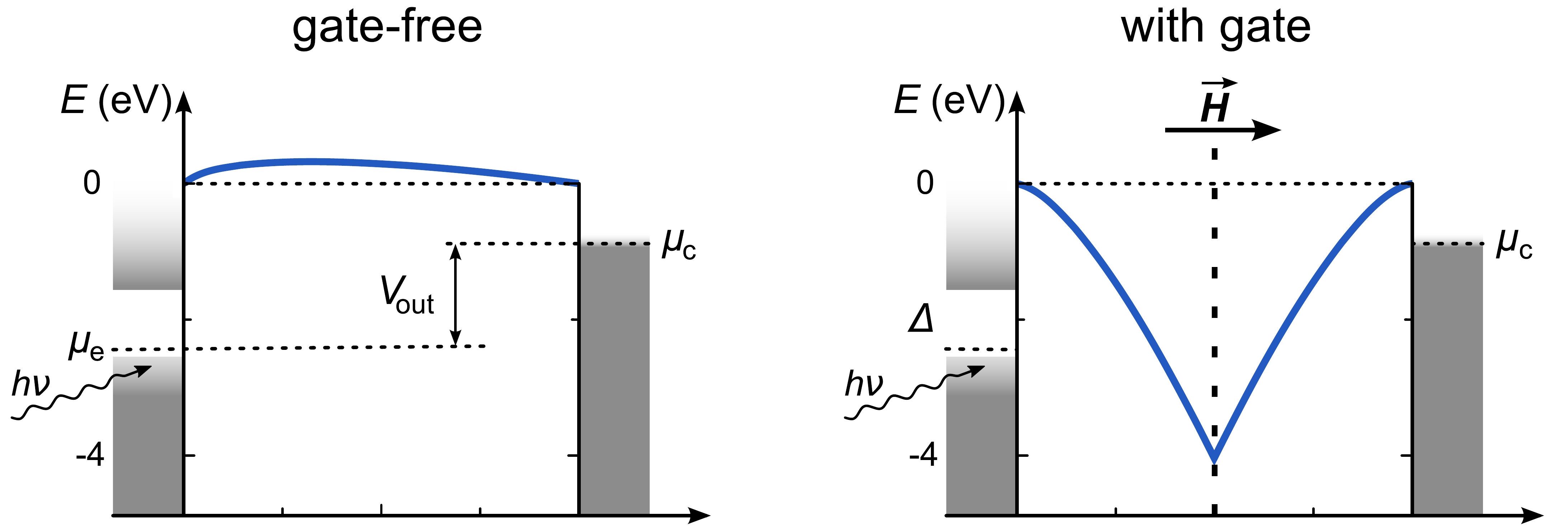}}
  \\
  \vspace*{0.02\textwidth}
  \subfloat[]{\hspace*{-.04\columnwidth}\includegraphics[width=1.05\columnwidth]{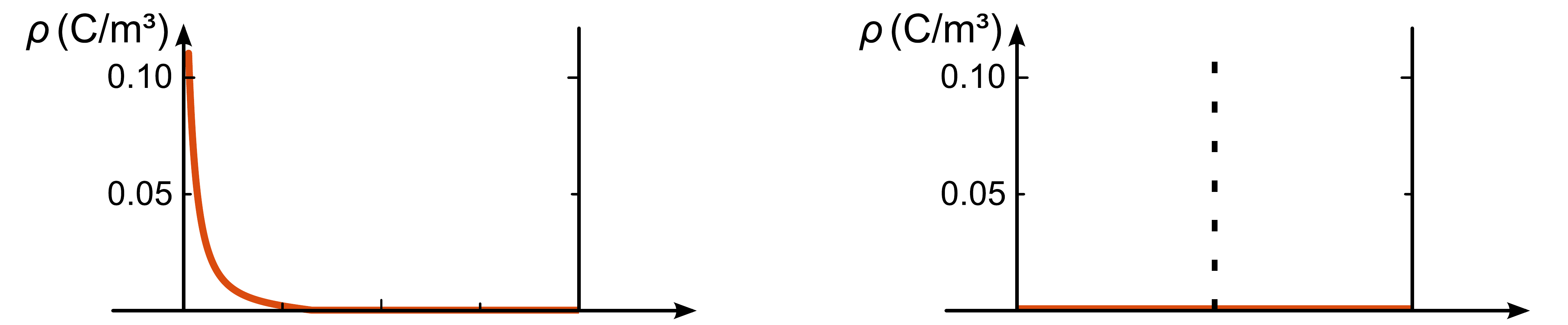}}
  \\  
  \vspace*{0.02\textwidth}
  \subfloat[]{\hspace*{-.04\columnwidth}\includegraphics[width=1.05\columnwidth]{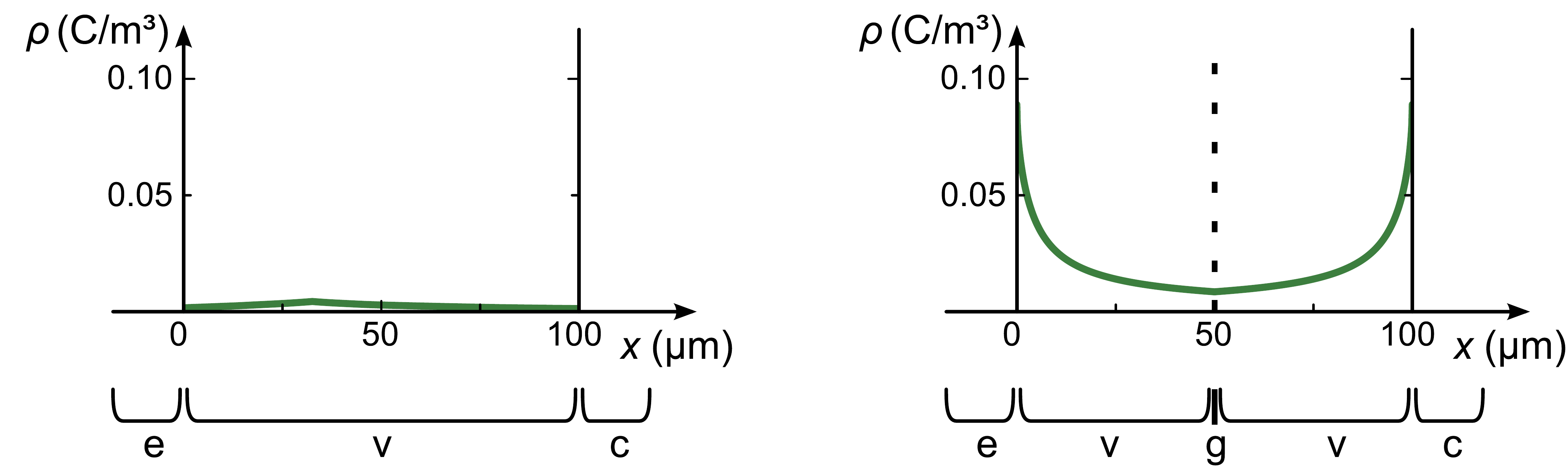}}
  \caption{Sketch of the working principle of thermoelectronic generators without (left) and with (right) a gate. The gate, positively biased with $\Vge=\myunit{6}{V}$, is mounted between emitter and collector; a homogeneous magnetic field is applied in x-direction. (a) Calculated potential profile. (b) Calculated density of electrons in the space-charge cloud. These electrons do not reach the collector. (c) Calculated density of electrons in the emitter--collector current. These electrons do reach the collector. The calculations and figures refer to the following parameters: $\phie=\myunit{2.5}{eV}$, $\phic=\myunit{0.9}{eV}$, $\Te=\myunit{1227}{\mycelsius}$ (\myunit{1500}{K}), $\Tc\leq\myunit{250}{\mycelsius}$, $\dec=\myunit{100}{\upmu m}$, $\Vout=(\phie-\phic)/e$, $w\rightarrow 0$. The labels ``\mue'' and ``\muc'' refer to the electrochemical potential of the emitter and collector; ``$h\nu$'' designates the incoming photons; ``c'', ``g'', ``e'', ``v'' denote the collector, gate, emitter, and vacuum locations, respectively. The data shown here were calculated using the 1D model (see Appendix~\ref{sec:model-calc:1D}).}
  \label{fig:1}
\end{figure}

Here we show that the space-charge problem can be solved in a plasma-free process. This process involves only electrons but no ions. It is therefore best characterized as ``thermoelectronic''. To remove the static space charges, a positively charged gate electrode is inserted into the emitter--collector space to create a potential trough. In a virtually lossless process this trough accelerates the electrons away from the emitter surface and decelerates them as they approach the collector (Fig.~1). A nominally homogeneous magnetic field $H$ applied along the electron trajectories prevents loss of the electrons to a gate current \Ig by directing them through holes in the gate on helical paths circling straight axes. This process turns the static space-charge cloud, which previously blocked the electron emission, into a useful output current (Fig.~1b,c). The design is analogous to that of ion thrusters used for spacecraft propulsion.

\begin{figure*}
  \subfloat[]{\includegraphics[width=0.27\textwidth]{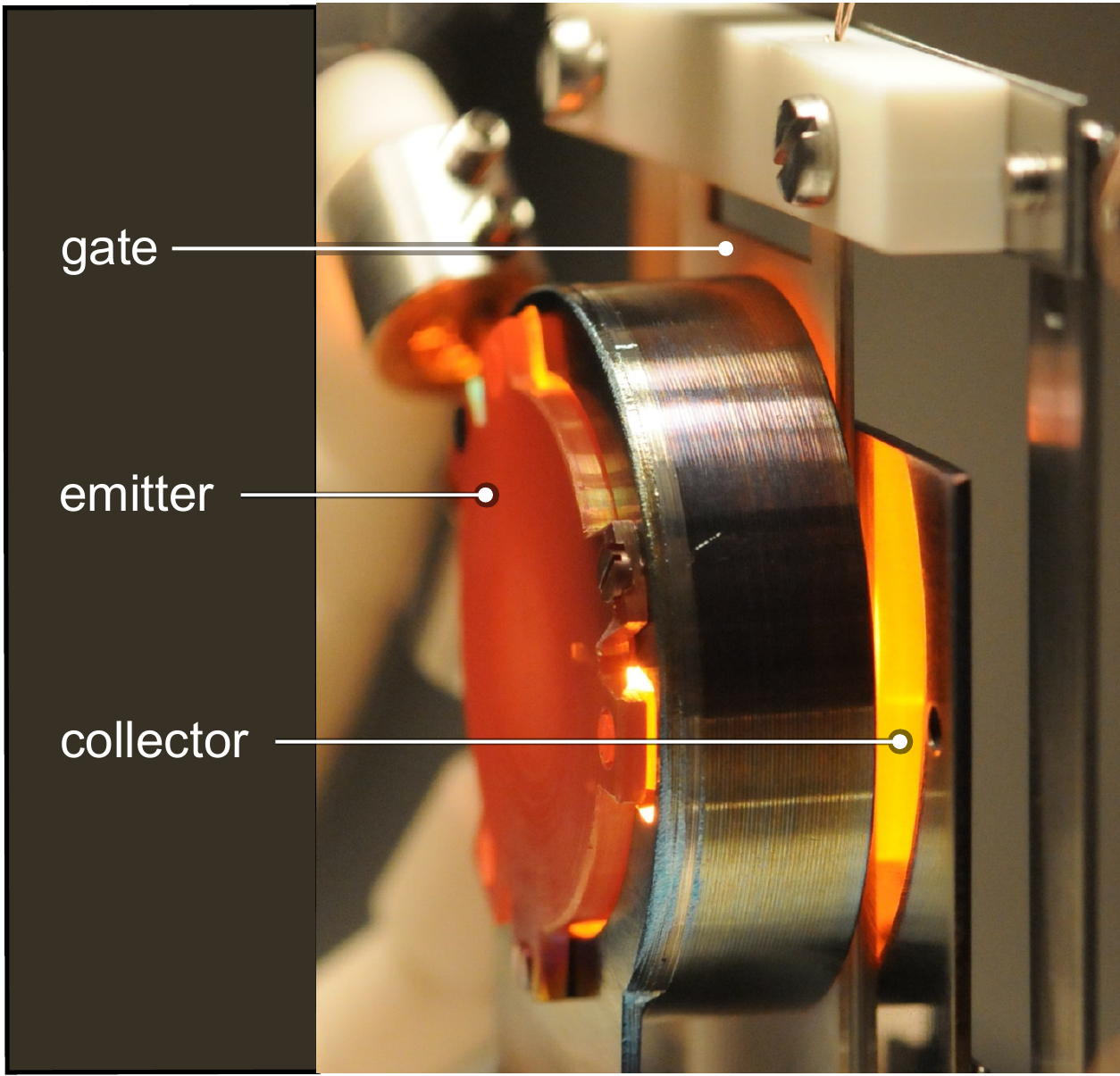}}
  \hspace{0.02\textwidth}
  \subfloat[]{\includegraphics[width=0.28\textwidth]{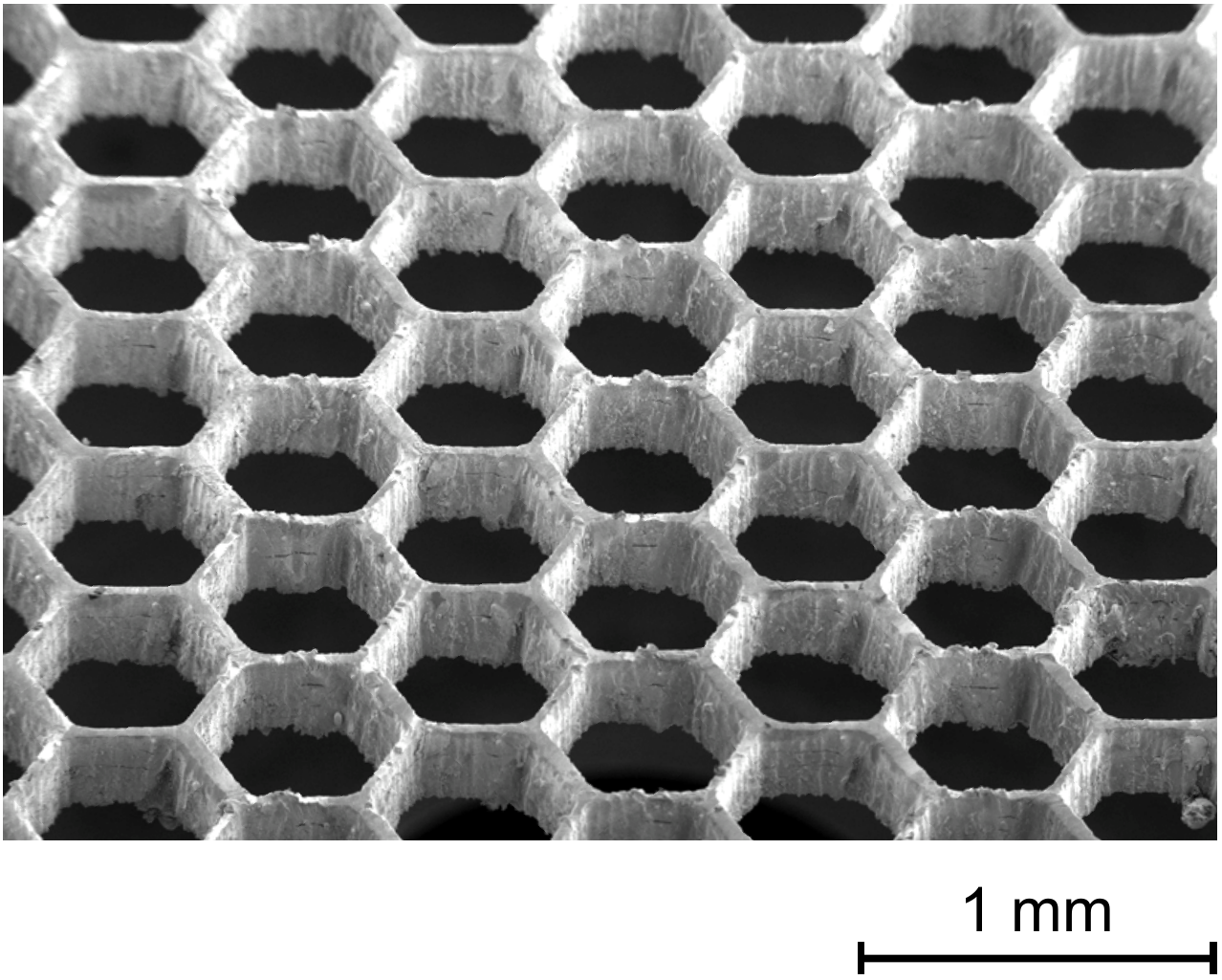}}
  \hspace{0.02\textwidth}
  \subfloat[]{\includegraphics[width=0.39\textwidth]{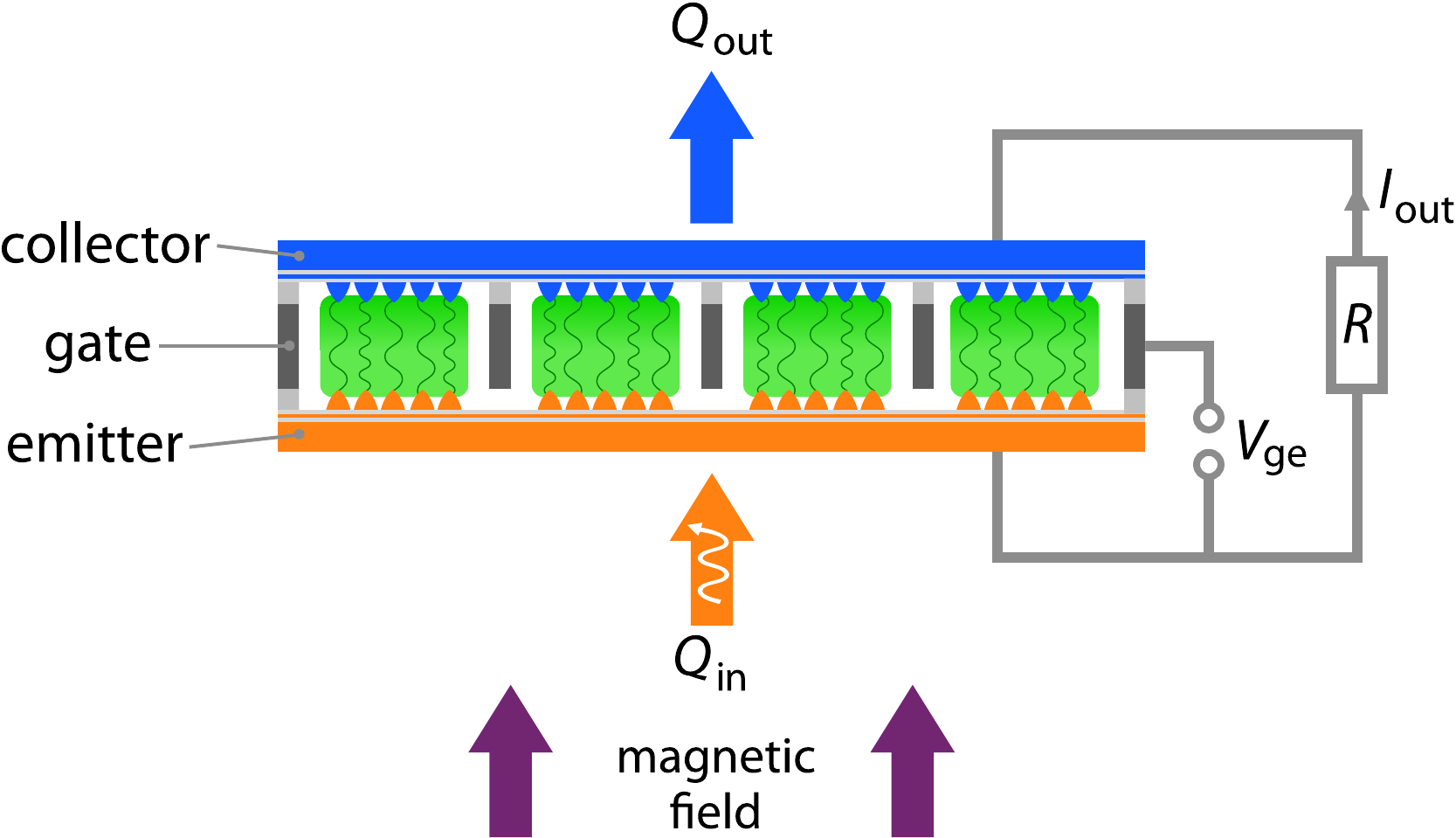}}
  \caption{(a) Photograph of a generator used in these experiments. The glowing orange disk (left) shows the back of the resistively heated emitter (BaO dispenser); the yellowish disk edge on the right shows the reflection of the glowing emitter on the collector surface (steel). (b) Micrograph of a grid (200-$\upmu\mathrm{m}$-thick tungsten foil, $w=\myunit{0.6}{mm}$) used as gate. (c) Setup of a possible microfabricated generator. The emitter and collector consist of wafers coated with heterostructures (gray lines) designed for the desired work function, thermal and infrared properties. The emitter and collector surfaces comprise nano-hillocks for local field enhancements. The green areas mark the regions of the electron flow through the vacuum, the direction of \Iout corresponds to the flow of positive charges.}
  \label{fig:2}
\end{figure*}

To investigate the effectiveness of the gate in removing the space charges, we fabricated a set of thermoelectronic generators as model systems (Figs.~2a,b; see Appendix~\ref{sec:setup}). The function of the generators was furthermore modeled by numerical calculations of the electron emission, space-charge formation and electron trajectories (see Appendix~\ref{sec:model-calc}). Experiment and model calculations provide consistent evidence that, by applying emitter-gate voltages of $\Vge \mytilde 2\text{--}\myunit{10}{V}$, the exact value being a function of the geometrical design of the generator, we can indeed remove the static space-charge clouds (Figs.~1b,c, 3a). The gate potential enables operation of the generators in vacuum with emitter--collector spacings of tens of micrometers (see Fig.~3b).

\begin{figure}
  \subfloat[]{\includegraphics[width=0.35\textwidth]{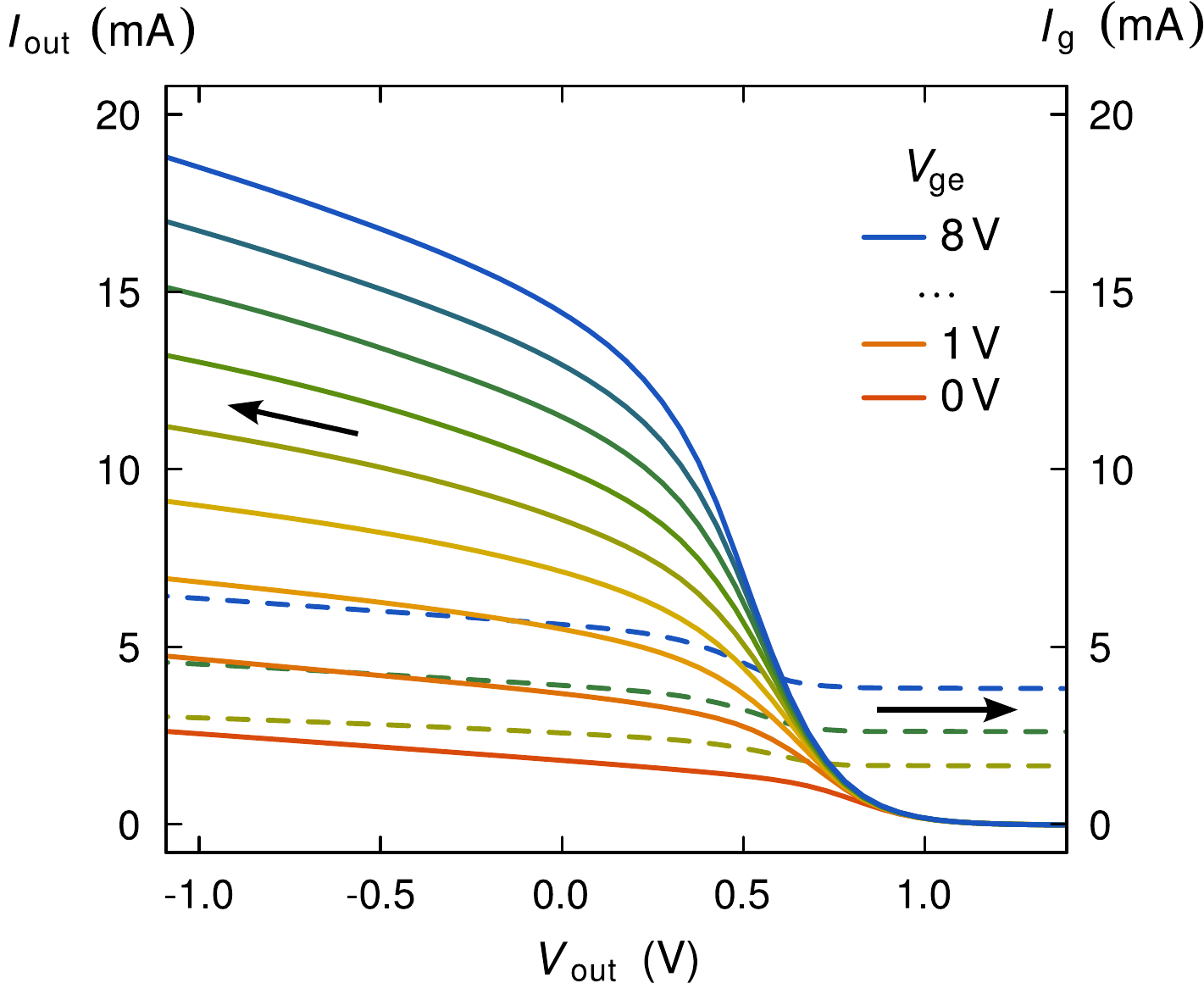}}
  \\
  \subfloat[]{\includegraphics[width=0.47\textwidth]{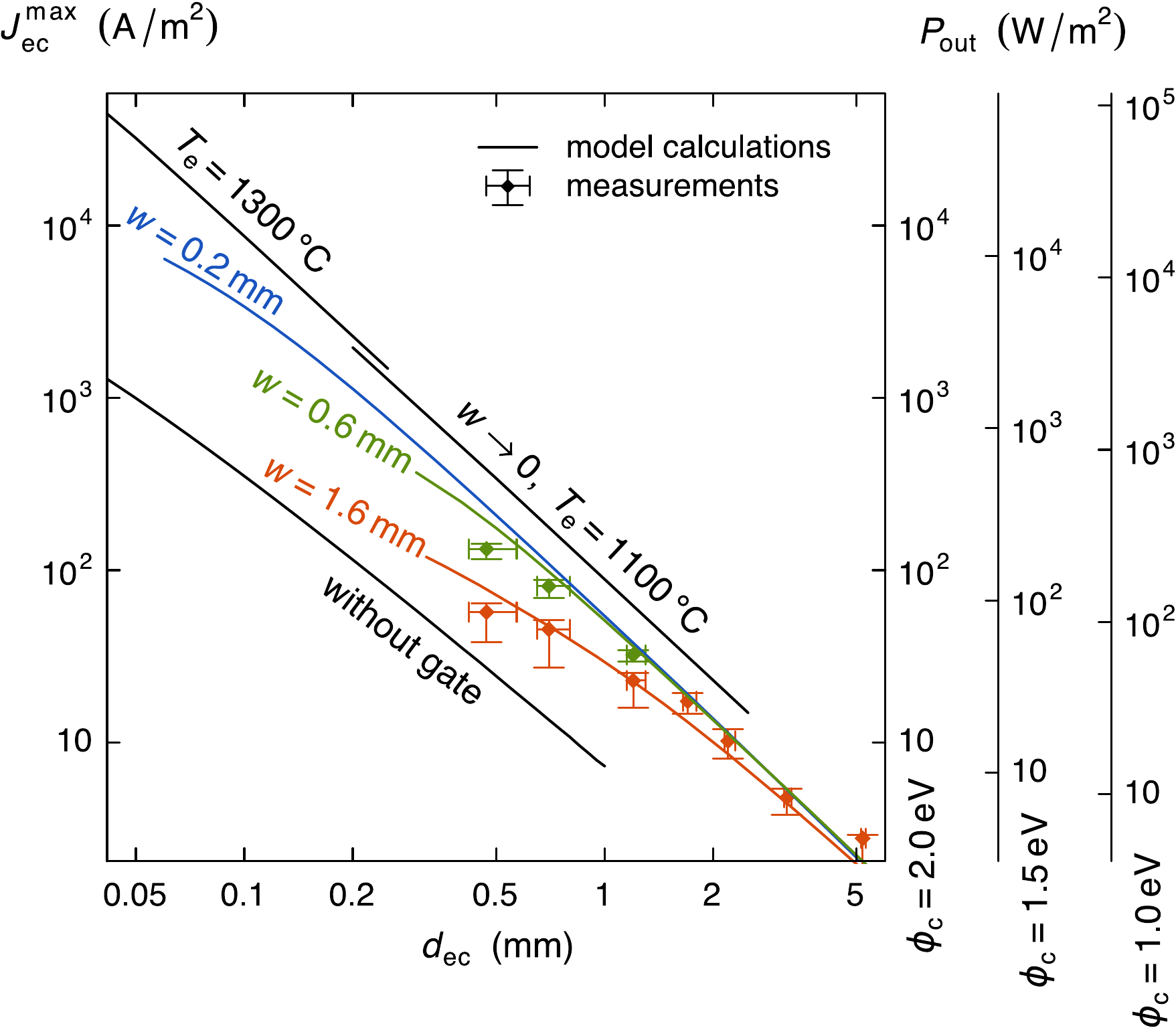}}
  \caption{(a) Output current and gate current measured as a function of \Vout for several gate voltages at $\Te=\myunit{1000}{\mycelsius}$, $\Tc=\myunit{500}{\mycelsius}$, $w=\myunit{1.6}{mm}$ and $\dec=\myunit{700}{\upmu m}$. Nominally identical BaO dispenser cathodes ($\phie \,\mytilde\, \phic \,\mytilde\, \myunit{2.2}{eV}$) were used for the emitter and collector. (b) Measured and calculated dependences of \Jecmax on \dec. The data was measured at $\Te=\myunit{1100}{\mycelsius}$, $\Tc\,\mytilde\,\myunit{500}{\mycelsius}$, $\Vge=\myunit{6}{V}$; the calculated current density refers to the density within the gate mesh. The output power densities \Pout were calculated from \Jecmax for $\phie=\myunit{3}{eV}$ using $\Pout=\Jecmax (\phie-\phic)/e$.
The error bars refer to the errors in determining \phie, \phic, and \dec. The data for $w \rightarrow 0$ and for the curve labeled ``without gate'' were calculated using the 1D model including the thermal distribution of electron velocities (see Appendix~\ref{sec:model-calc:1D}); the data for $w > 0$ were calculated using the quasi-3D model (see Appendix~\ref{sec:model-calc:3D}).}
  \label{fig:3}
\end{figure}

Although, as will be shown below, the generators operate with high efficiencies at large \dec, the value of the emitter--collector current \Iec decreases with \dec. This is illustrated by Fig.~3b, which shows that the density \Jecmax of the emitter--collector current at which the maximal output power is obtained, \Iecmax, scales for large \dec with $1/\dec^2$. At small \dec, \Jecmax approaches the current density of gate-free generators, because the electric field becomes small inside the mesh holes if $\dec \ll w$, where $w$ is the grid-mesh diameter defined for hexagonal grids as the distance between opposite corners. For grids with finite conductor widths, \Jecmax is furthermore reduced because for $t < 1$, a fraction of the emitted current is lost to \Ig. Here, $t$ is the gate transparency, the fraction of the gate area not covered by the conductor. This effect can be minimized by optimizing the gate geometry and by inducing an inhomogeneous electron emission, for example by using nanotubes grown on the emitter. In the latter case, \Jecmax may be increased further by gate-field-enhanced emission.

\begin{figure*}[hbpt]
\includegraphics[width=1.8\columnwidth]{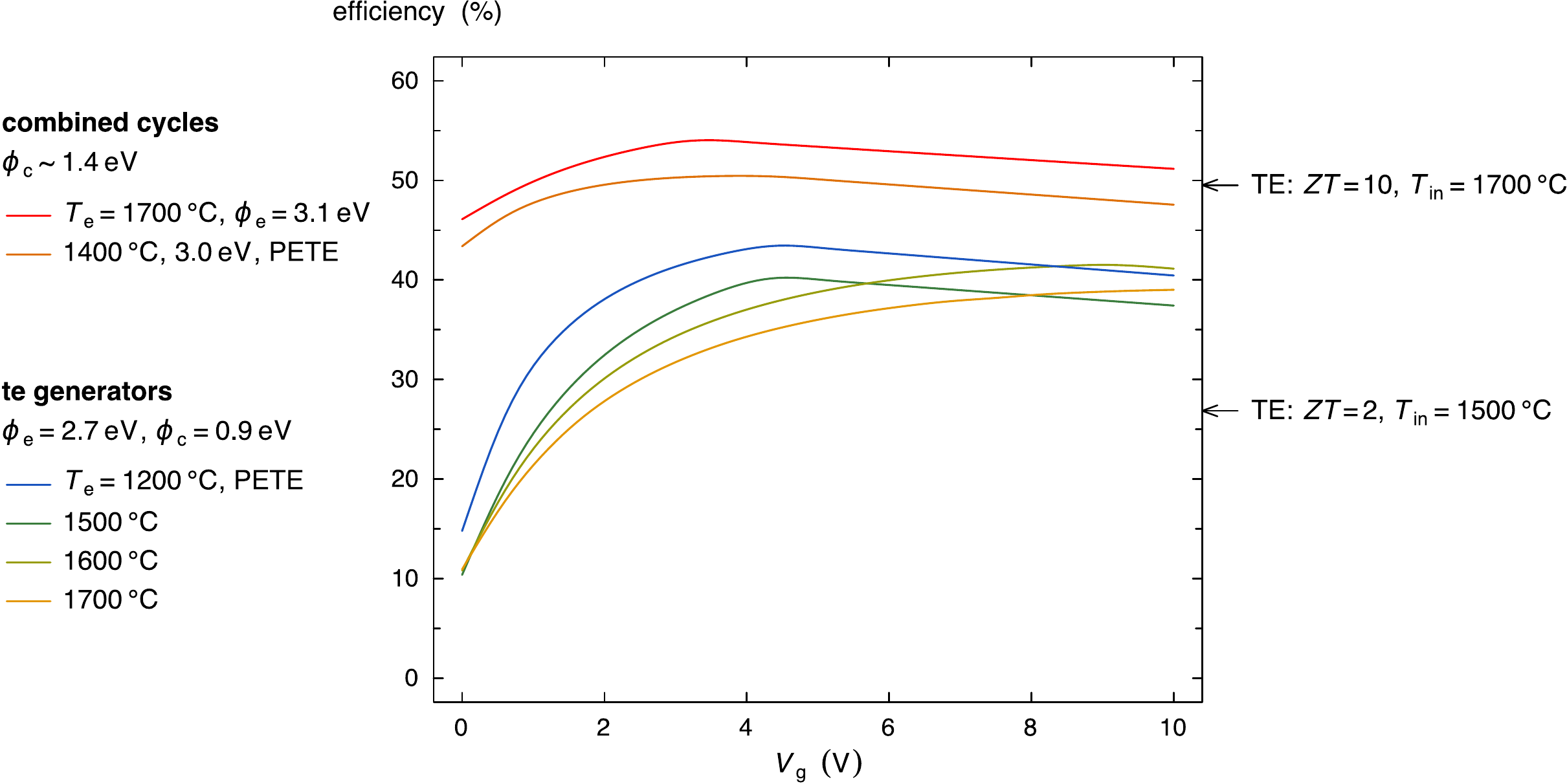}
\caption{Heat--to--electric-power conversion efficiencies calculated as a function of the gate voltage of stand-alone thermoelectronic generators working at a series of emitter temperatures ($\Tc=\myunit{200}{\mycelsius}$) and of systems comprising a thermoelectronic generator as topping cycle ($\dec=\myunit{30}{\upmu m}$). In the combined-cycle systems, the thermoelectronic generators operate between \Te and $\Ts=\myunit{600}{\mycelsius}$. The work functions were selected for optimal performance and $\Te=\myunit{1700}{\mycelsius}$ to allow a comparison with the efficiency given for the stand-alone system. State-of-the-art steam turbines were presumed to work as bottom cycle, receiving heat at \Ts and converting it into electricity with $\eta=\myunit{45}{\%}$. Owing to the high \Tout of the thermoelectronic generator, \phie and \phic can have rather large values. For the calculation of the efficiencies of the thermoelectronic PETE analogue, a band gap of \myunit{1.5}{eV} and electron affinities of 1.6 and \myunit{1.85}{eV} were considered for the stand-alone and the combined-cycle systems, respectively (see Appendix~\ref{sec:model-calc:eff}). Light--to--electric-power conversion efficiencies for a light-concentration of 5000 are shown for the PETE systems.
The image also lists the efficiencies of hypothetical thermoelectric generators with figures of merit of $ZT=2$ and $10$ at temperatures between \Tin and \myunit{200}{\mycelsius} (see~\cite{bell_cooling_2008} and Appendix~\ref{sec:model-calc:eff}). For comparison, the maximum efficiency of single-junction solar cells is $\mytilde\myunit{34}{\%}$ (Shockley--Queisser limit~\cite{shockley_detailed_1961}) and the best research multi-junction photovoltaic cells have efficiencies of $\mytilde\myunit{43.5}{\%}$~\cite{green_solar_2012}.}
\label{fig:4}
\end{figure*}

Having confirmed that the space-charge cloud has been removed, we now explore the efficiency $\eta=\Pout/\Pin$, with which these generators transform heat into electric power. The output power of the generator, $\Pout=\Iout\Vout$, is maximal for $\Vout=(\phie-\phic)/e$, where \phie and \phic are the work functions
of the emitter and the collector~\footnote{Following common usage, (e.g., \cite{lang_theory_1971}), we define the work function of a material as the
energy required to move an electron with an energy equaling the chemical potential
from inside the material to a location far away from the surface.}, respectively, and $e$ is the elementary charge. For larger \Vout, some of the electrons lack the energy to reach the collector, whereas \Iec is independent of \Vout for smaller \Vout. We start to identify the efficiency limit by considering a simplified, ideal case, in which the input power \Pin is converted completely into an emitter--collector current consisting only
of electrons at the vacuum potential ($E=0$). If the electrons are only thermally emitted, the requirement that the back-emission current from the collector is so small that \Iecmax is positive entails that $\eta<1-\frac{\Tc}{\Te}$ (see Appendix~\ref{sec:model-calc:eff}). To generate this ideal current, a power of $\Pin=\Iecmax \phie / e$ is required. Therefore, $\etamax=1-\frac{\phic}{\phie}$ is a strict upper limit for the heat--to--electric power conversion efficiency. This limit also applies to devices in which the photoelectric effect is used.

In real devices, $\eta$ is reduced by several loss channels, which include the above-neglected thermal
energy carried from the emitter by \Iecmax, losses due to a finite \Ig, radiation losses from the emitter,
thermal conduction of the wires contacting the electrodes, and ohmic losses. Nevertheless, only the loss by the electron heat current causes a fundamental bound for the efficiency; the other loss effects can in principle be reduced to very small values.

Figure~4 shows the results of the model calculations of the generator efficiencies as a function of the gate voltage, considering the above-mentioned losses (see Appendix~\ref{sec:model-calc:eff}). Starting at $\Vge=0$, $\eta$ increases with \Vge as the gate potential sweeps the space charges into the collector. This increase demonstrates the usefulness of the gate field. At higher \Vge, $\eta$ decreases because the space charges have been removed and \Vge does not enhance \Iecmax beyond the maximum emission current, but increases the power $\Ig\Vge$ lost at the gate. For a given \phie, $\eta$ increases with increasing \Te due to higher emission currents until thermal radiation losses dominate. For the parameter range considered realistic for applications (\textit{e.g.}, $\dec=\myunit{30}{\upmu m}$, $t=0.98$, $\phic=\myunit{0.9}{eV}$~\cite{koeck_thermionic_2009}), maximum efficiencies of $\mytilde \myunit{42}{\%}$ are predicted. The calculated efficiencies (Figs.~4) are consistent with previous calculations of efficiencies of thermionic generators that were presumed to be devoid of space charges~\cite{houston_theoretical_1959, ingold_calculation_1961, lee_optimal_2012, moyzhes_thermionic_2005, rasor_thermionic_1991}. They compare well with those of photovoltaic solar cells~\cite{green_solar_2012}, thermoelectric materials~\cite{bell_cooling_2008, snyder_complex_2008}, and focused solar mechanical generators~\cite{mueller-steinhagen_concentrating_2004, richter_solarpaces_2010}. The results on combined cycles shown in Fig.~4 reveal that by using thermoelectronic converters as topping cycles the efficiency of state-of-the-art coal combustion plants may be increased from \myunit{45}{\%} to \myunit{54}{\%}, corresponding to a reduction of emissions such as CO$_2$ by $\mytilde\myunit{17}{\%}$.

\section{Conclusion}

Optimization of the conversion efficiencies requires the development of metal or semiconductor surfaces with the desired effective work functions and electron affinities, respectively, which may also be done by nanostructuring the electrode surfaces. These surfaces need to be stable at high temperatures in vacuum. The tunability of the gate field opens possibilities to alter the converter parameters during operation. Although the need to generate Cs$^+$ ions to neutralize the space-charge cloud is eliminated, adatoms of elements such as Cs can be used to lower the work function of the electrodes, in particular of the collector. For high efficiency, the devices must be thermally optimized to minimize heat losses through the wiring. Furthermore, thermal radiation of the emitter must be reflected efficiently onto the electrode. For ballistic electron transport between emitter and collector, a vacuum of better than \myunit{0.1}{mbar} is also required, reminiscent of radio tubes.

Such devices may be realized, for example, in a flip-chip arrangement of oxide-coated wafers separated by tens of micrometers using thermal-insulation spacers as sketched in Fig.~2c. This produces hundreds of Watts of power from active areas of some \myunit{100}{cm^2}. The magnetic fields, typically $\leq\myunit{1}{T}$ with large tolerances in strength and spatial distribution, can be generated by permanent magnets or, for applications such as power plants, by superconducting coils. Achieving viable, highly efficient devices requires substantial further materials science efforts to develop the functional, possibly nanostructured materials, as well as engineering efforts to achieve a stable vacuum environment in order to minimize radiative and conductive heat losses, and to ensure competitive costs. Remarkably, however, no obstacles of a fundamental nature appear to impede highly efficient power generation based on thermoelectronic energy converters.

\begin{acknowledgments}
The authors gratefully acknowledge discussions with H.\,Boschker, R.\,Kneer, T.\,Kopp, H.\,Queisser, A.\,Reller, H.\,Ruder, A.\,Schmehl, and J.\,Weis as well as technical support by B.\,Fenk and A.\,Herrnberger. One of us (THG) would like to acknowledge informative conversations on the use of triodes with longitudinal magnetic fields to generate Cs plasmas in thermionic generation with the late Boris Moyzhes, and also acknowledges support for part of the work at Stanford by the U.S. Department of Energy, under contract DE-DE-AC02-76SF00515.
\end{acknowledgments}

\clearpage

\appendix

\section{Experimental Setup and Procedures}
\label{sec:setup}

In the model systems the electrodes were mechanically mounted in a vacuum chamber (base pressure \myunit{10^{-7}}{mbar}) to facilitate the study of various converter configurations. As emitters, commercial, resistively heated BaO-dispenser cathodes~\cite{heatwave_labs_inc.__????} with a temperature-dependent work function in the range $\myunit{2.0}{eV} < \phie < \myunit{2.5}{eV}$ and an emitting area of \myunit{2.8}{cm^2} were used. The gates were laser-cut tungsten foils, the spacers aluminum oxide foils, and the collectors either consisted of polished steel plates or were BaO-dispenser cathodes. The collector work functions were determined from the $\Iout(\Vout)$-characteristics and additionally from the Richardson-Dushman saturation current.
The emitters are ohmically heated, \Te was measured with a pyrometer. The magnetic field is generated by two stacks of NdFeB permanent magnets mounted on both sides of the emitter-gate-collector assembly. They created at the gates $\mytilde (200 \pm \myunit{10)}{mT}$. Photon-enhancement of the emission was not applied.  Electrical measurements were performed with source-measurement units (Keithley 2400) in 4-wire sensing.

\section{Model Calculations}
\label{sec:model-calc}

\subsection{One-dimensional models}
\label{sec:model-calc:1D}

For the calculations of the current densities in gate-free, plane-parallel configurations the one-dimensional space-charge theory of Langmuir~\cite{langmuir_effect_1923} and Hatsopoulos~\cite{hatsopoulos_thermionic_1979} was used to determine the space-charge potential.
To incorporate the effect of the gate in the one-dimensional approach, these models were extended to include the potential generated by an idealized gate, assumed to be a metal plate that is transparent for electrons and to create a homogeneous electric field.  
Calculations of the electric field of a patterned metal grid with the commercial electric field solver COULOMB~\cite{integrated_engineering_software_ies._2011} showed that for $\dec > w$ the generated field is virtually identical to the field of an idealized gate. The 3D calculations of the electric field distribution and the electron paths in the electric gate field and the applied magnetic field done with the commercial software LORENTZ~\cite{integrated_engineering_software_ies._2011} showed that the electrons are forced on quasi-one dimensional paths by the magnetic field and are thus channeled through the gate openings.

To explore \Jemax as a function of \Vge below we calculate the course of the electric potential in the vacuum gap. For this we consider a symmetrical setup, the gate being located in the middle between emitter and collector. The gate potential for electrons is given by 

\begin{align*}
\phig(x) &= - \frac{2\Vge}{\dec} x \quad \mathrm{for}\; 0 \leq x \leq \frac{\dec}{2},
\intertext{and}
\phig(x) &= - \frac{2\Vge}{\dec} (\dec-x) \quad \mathrm{for}\; \frac{\dec}{2} \leq x \leq \dec.
\end{align*}

At maximum power output, emitter and collector have the same local vacuum potential. We assume the collector to be cold enough that back emission is negligible, as discussed in Ref.~\cite{hatsopoulos_thermionic_1979}.

If the thermally distributed initial velocity of emitted electrons is neglected, the Poisson equation is given by
\begin{align*}
\Delta \Psi(x) = - \frac{J}{\epsilon_0} \left( - \frac{2e}{\me} \Psi(x) \right)^{-1/2},
\end{align*}
where $\Psi(x)$ is the total electrostatic potential for negative charges, consisting of the contribution of the gate and the space-charge potential. This equation is solved analytically, analogous to the Child-Langmuir law~\cite{langmuir_effect_1913,child_discharge_1911}, yielding

\begin{align}
J = \epsilon_0 \sqrt{\frac{e}{6\me}} \frac{\Vge^{3/2}}{\dec^2}. \label{eq:1}
\end{align}

Remarkably, the current density shows the same behavior $J \propto V^{3/2} / d^2$ as the Child-Langmuir law. 

If the thermal velocity distribution is included, the Poisson equation becomes
\begin{align*}
\Delta \Psi(x) & = - \frac{en_0}{\epsilon_0} \exp \left[ - \frac{e}{\kB T} \Psi(x) \right] \cdot \\ 
& \cdot \left\lbrace 1 \pm \operatorname{erf} \left[ \frac{e}{\kB T} \left( \Psimax - \Psi(x) \right) \right] \right\rbrace,
\end{align*}
where $en_0$ is the space-charge density at the emitter surface and \Psimax the maximum of the space-charge potential in the inter-electrode space. The plus sign is valid for $x \leq x_\mathrm{max}$, the minus sign for $x \geq x_\mathrm{max}$, with $x_\mathrm{max}$ being the position of $\Psi_\mathrm{max}$. $n_0$ can be determined from the Richardson-Dushman equation~\cite{hatsopoulos_thermionic_1979}; it is a function of \phie and \Te. This self-consistent differential equation has to be solved numerically.

We used Mathematica 8.0 for the numerical calculations. For each iteration step, the change of the space-charge potential has to be kept small, as already a small modification of $\Psi(x)$ can lead to a strong modification or even a divergence of the solution. Therefore, the solution has to be approached slowly to impede a strong oscillatory behavior.

The model calculations labeled “$w \rightarrow 0$” in Fig. 3b were obtained using the ideal transparent gate model including the electron velocity distribution. 

\subsection{The quasi-3-dimensional current tube model}
\label{sec:model-calc:3D}

To take the inhomogeneities of the electric field of the gate electrode into account, the interelectrode space was subdivided into narrow prisms, which extend from the emitter to the collector surface. We calculated the average gate potential in each prism with the electric field solver COULOMB~\cite{integrated_engineering_software_ies._2011}. We apply a linear regression to determine the mean electric field, which can be used in the one-dimensional gate model. We then calculate the current density for each prism separately. Thereby the interactions between the prisms were neglected, which is a good approximation for the case of small inhomogeneities in the space-charge density. The total current density  was obtained by summing up the contributions from all tubes.

Due to the high computational effort required to solve the 1D model including the thermally distributed initial electron velocity, the analytical solution (Eq.~\ref{eq:1}) was used to determine the current density, which yields a good approximation in the voltage range considered. However, it does not account for the temperature-dependence of the current density. 

\subsection{Efficiency calculations}
\label{sec:model-calc:eff}

\subsubsection*{Calculation of the ultimate efficiency limit}
The Richardson-Dushman equation describes the current density for electrons emitted from a metal surface~\cite{ashcroft_solid_1976}. It is obtained by using the equation $J=-nev$ and integrating the Fermi distribution \fFD over all electrons with a positive velocity normal to the emitting surface, \textit{i.e.},

\begin{align}
\JRD &= - e \iiint\limits_{v_\mathrm{x}>0} \upd \vec{v} v_\mathrm{x} f_\mathrm{FD}(\vec v) \approx \frac{e\me}{4\pi^2 \hbar^3} \exp \left( \frac{-\phi}{\kB T} \right) \cdot \notag \\
& \cdot \int\limits_0^\infty \upd v_x \int\limits_{-\infty}^\infty \upd v_y \int\limits_{-\infty}^\infty \upd v_z v_x \exp \left( \frac{-\me v^2}{2\kB T} \right) =  \notag\\
 &= - \ARD T^2 \exp \left( \frac{-\phi}{\kB T} \label{eq:2}\right).
\end{align}
\ARD: Richardson-Dushman constant, $\phi$: work function, $T$: surface temperature, $v$: electron velocity.

If all non-fundamental channels of heat loss are neglected, heat is lost from the emitter only by the transport of electrons. This electron cooling \Pel is given by~\cite{hatsopoulos_thermionic_1979}

\begin{align}
\Pel &= \int\limits_0^\infty \upd v_x \int\limits_{-\infty}^\infty \upd v_y \int\limits_{-\infty}^{\infty} \upd v_z v_x \left( \phi + \frac{\me v^2}{2} \right) \fFD(v) = \notag \\
& = \frac{\JRD}{e} (2\kB \Te + \phie). \label{eq:3}
\end{align}

Assuming there is no space-charge cloud limiting the transfer of electrons across the vacuum gap, both \Jemax and the back-emission \Jbemax from the collector are given by the respective Richardson-Dushman current densities (Eq.~\ref{eq:2}). 

Taking into account the heat transported back to the emitter by the back-emission, the efficiency is obtained to be:

\begin{align}
\eta = \frac{\Jecmax(\phie-\phic)}{\Jemax(\phie + 2\kB \Te)-\Jbemax(\phie + 2\kB \Tc)}. \label{eq:4}
\end{align}

This value is known to always be smaller than the Carnot efficiency~\cite{baksht_thermionic_1978,hatsopoulos_thermionic_1973}.

However, the efficiency may be ultimately increased if electrons are emitted only at a discrete energy \Enull, so that the $2\kB T$-terms in Eqs.~\ref{eq:3} and~\ref{eq:4} disappear. For this case, however, the Richardson-Dushman equation does not apply. Instead, the emitted current density has to be calculated for a hypothetical material with the discrete energy level \Enull, from which the emission of electrons occurs. This level may be at or above the vacuum level \Evac. This calculation can be performed by inserting a $\delta$-function to describe the discrete density-of-states at $E=\Enull$. In this case, in Eq.~\ref{eq:2} no Gaussian-integral has to be determined and the resulting, discrete current density \JEnull does not have a term with coefficient $T^2$.

As for any thermoelectronic generator, an output power is only generated for 
\begin{align*}
\Jecmax = \Jemax - \Jbemax = \JeEnullmax - \JbeEnullmax > 0,
\end{align*}
implying
\begin{align*}
\exp \left( \frac{-\phie}{\kB\Te}\right) - \exp \left( \frac{-\phic}{\kB\Tc}\right) > 0.
\end{align*}
It follows
\begin{align*}
\frac{\phic}{\phie} > \frac{\Tc}{\Te},
\end{align*}
and therefore 
\begin{align*}
\eta &= \frac{\Jecmax/e\cdot(\phie-\phic)}{\Jecmax/e\cdot\phie} = \frac{\phie-\phic}{\phie} = \\
&=  1-\frac{\phic}{\phie} < 1-\frac{\Tc}{\Te} = \eta_\mathrm{Carnot}.
\end{align*}
For  $J \rightarrow 0$, it follows
\begin{align*}
\frac{\phic}{\phie} \rightarrow \frac{\Tc}{\Te},
\end{align*}
and consequently:
\begin{align*}
\eta \rightarrow \eta_\mathrm{Carnot}.
\end{align*}
 
As can be seen, the efficiency approaches the Carnot limit if the net current across the vacuum gap approaches zero, \textit{i.e.}, if the system approaches equilibrium. Consequently, the output power approaches zero when the efficiency approaches the Carnot limit. This is a very typical behavior for any realistic heat engine (see, \textit{e.g.}, Refs.~\cite{curzon_efficiency_1975, fuchs_dynamics_1996}).

\subsubsection*{Stand-alone generators}

To calculate the efficiency of realistic thermoelectronic generators, the calculations presented in Refs.~\cite{houston_theoretical_1959,ingold_calculation_1961,rasor_thermionic_1991} were extended to include both the gate energy loss and the dependence of \Iemax on the gate voltage. In determining the generator efficiency, the power \Pg required to sustain the gate electric field is subtracted from the output power:

\begin{align*}
\eta = \frac{\Pout-\Pg}{\Pin},
\end{align*}
where \Pin is the heat input and \Pout the power delivered to the load. It is given by

\begin{align*}
\Pout = \left( \frac{\phie-\phic}{e} - \Vlead \right) \Iecmax,
\end{align*}
with the net current flowing to the collector
\begin{align*}
\Iecmax = t \Iemax - \Ibemax,
\end{align*}
and the voltage drop in the leads connecting the load with the emitter (\Rle) and collector (\Rlc)
\begin{align*}
\Vlead = \Iecmax\Rlc + \left( \Iemax - t\Ibemax\right) \Rle.
\end{align*}
Here, \Iemax is the space-charge limited current emitted from the emitter, which is calculated from the models described above and \Ibemax the back-emission current emerging from the collector. It has to be considered that \Ibemax is also reduced by the space-charge potential. Therefore, it is given by
\begin{align*}
\Ibemax = \IRD \exp \left( -\frac{\Psimax}{\kB\Tc}\right),
\end{align*}
with the Richardson-Dushman current \IRD and the maximum of the inter-electrode potential \Psimax. 

In the steady state the heat input equals the sum of all channels of heat loss from the emitter:
\begin{align*}
\Pin = \Pel + \Prad + \Pcond,
\end{align*}
with the electron cooling:
\begin{align}
\Pel &= \frac{\Iemax}{e} (\phie + \Psimax + 2\kB\Te) - \notag \\
& - \frac{t\Ibemax}{e}(\phie+\Psimax+2\kB\Tc), \label{eq:5}
\end{align}
the radiation cooling:
\begin{align*}
\Prad = \sigma\epsilon A(\Te^4-t\Tc^4),
\end{align*}
($A$: emitter area, $\sigma$: Stefan-Boltzmann constant, $\epsilon \sim 0.1$: effective emissivity of the electrode system~\cite{rasor_thermionic_1991}) and the heat conduction across the emitter lead:
\begin{align*}
\Pcond = \frac{L}{2\Rle}(\Te-\Tnull)^2 - \frac{\Rle}{2}(\Iemax-t\Ibemax)^2,
\end{align*}
where the lead is assumed to be metallic and to follow the Wiedemann-Franz law. With the Lorentz number $L$ the thermal conductivity can consequently be expressed as $LT_\mathrm{mean}/\Rle$. The load is assumed to be at ambient temperature \Tnull. The second term in this equation arises from half of the Joule heat produced in the lead effectively being transported to the emitter, which can be shown by solving the heat flow equation~\cite{ingold_calculation_1961}.

\subsubsection*{Combined-cycle system}

In combined cycle systems the heat rejected by the collector (\Prej) is used to drive a secondary heat engine working at an efficiency of \etas. The power $\etas \Prej$ produced by this engine is added to the total produced power, hence
\begin{align*}
\etacc = \frac{\Pout - \Pg + \etas\Prej}{\Pin}.
\end{align*}

In the steady state \Prej is equivalent to the heat transported to the collector, given by the sum of an electronic, radiation, and conduction term 
\begin{align*}
\Prej = \Pelc + \Pradc + \Pcondc,
\end{align*}
where
\begin{align}
\Pelc &= \frac{t\Iemax}{e}(\phic+\Psimax+2\kB\Te) - \notag \\
&- \frac{\Ibemax}{e}(\phic+\Psimax+2\kB\Tc), \label{eq:6} \\
\Pradc &= \sigma\epsilon A(t\Te^4-\Tc^4), \notag
\end{align}
and
\begin{align*}
\Pcondc = \frac{L}{2\Rlc}(\Tc-\Tnull)^2 - \frac{(\Iecmax)^2\Rlc}{2}.
\end{align*}

\subsubsection*{Losses specific to solar heating}

For solar heated thermoelectronic generators another fundamental channel for heat loss arises which we take into account: to couple solar radiation into the emitter, the emitter needs to provide a highly absorbing surface \Ab (here ``b'' stands for black). This surface \Ab has a high emissivity and therefore emits a thermal power \Pb. The resulting, reduced light--to--electricity efficiency \etal is expressed in terms of the heat--to--electricity efficiency $\eta$:
\begin{align*}
\etal = (1-\frac{\sigma\Te^4}{c\Inull})\eta,
\end{align*}
where $c$ is the concentration-factor of the incoming solar radiation onto the absorbing spot on the emitter~\footnote{$c$ is not to be confused with the effective concentration \ceff that is relevant in the context of PETE. It is $c\Ab=\ceff A$.} and \Inull the intensity of the incoming solar radiation.

\subsubsection*{PETE-efficiencies}

To calculate the efficiency of a PETE device incorporating a gate electrode, we first assume a given emitted current density \Jepete and emitter temperature \Te. The latter is chosen such that the hypothetical Richardson-Dushman current density across the electron-affinity barrier (\Ea) is at least 100 times larger than \Jepete. For an ideal PETE-device we then expect an electron yield of 1 electron per above-bandgap-photon~\cite{schwede_photon-enhanced_2010}, as photoexcited electrons can then be assumed to be thermally emitted significantly faster than they recombine.

From \Jepete, which defines the emission capability of the emitter, we then calculate the space-charge limited current density \Jecmax from the 1D model described above (taking into account the thermally distributed starting velocity of the electrons). This defines the input power actually required to maintain a stable emitter temperature and, consequently, the required incident light concentration \ceff. For the data shown, this typically yields $\ceff \,\mytilde\, 500$. To satisfy the self-consistency, from \ceff and \Jecmax we finally calculate the bandgap \Eg that yields the required rate of photoexcitations into the conduction band.

We assume the chemical potential to be in the middle between the worst case (middle of the bandgap) and the best case (bottom of the bandgap). Consequently, the emitter work-function is
\begin{align*}
\phie=\Ea + \frac{3}{4}\Eg.
\end{align*}

From \phie and \Jecmax the efficiencies of both stand-alone and combined-cycle PETE devices can be calculated as described above.

\subsubsection*{Intrinsic electronic heat losses}

Below, the relative importance of the channels of heat loss will be discussed for the peak of the efficiency of the \myunit{1600}{\mycelsius}-curve shown in Fig.~4. Although the resulting numbers may slightly vary for other configurations, the ratios of the different contributions remain essentially the same.

At the peak of the efficiency of the \myunit{1600}{\mycelsius}-curve shown in Fig.~4 the total input power of $\Pin=\myunit{78.1}{W/cm^2}$ is mainly consumed by the electron cooling of $\Pel=\myunit{67.3}{W/cm^2}$. Therefrom, \myunit{60.0}{W/cm^2} are consumed by the emitted electrons to overcome \phie and \myunit{7.3}{W/cm^2} arise from the thermally distributed electron velocity (the $2\kB T$-terms in Eqs.~\ref{eq:5} and~\ref{eq:6}). The remaining loss splits up between thermal radiation ($\Prad=\myunit{7.0}{W/cm^2}$) and conduction across the lead wires ($\Pcond=\myunit{3.8}{W/cm^2}$). In this configuration the system delivers a power of $\Pout=\myunit{36.4}{W/cm^2}$ to the load cycle, while $\Pg=\myunit{4.0}{W/cm^2}$ are consumed on the gate. The resulting net output power of \myunit{32.4}{W/cm^2} corresponds to an efficiency of $\eta=\myunit{42}{\%}$.

\subsubsection*{Efficiency of thermoelectric generators}

For comparison, efficiencies of hypothetical thermoelectric generators are given in Fig.~4. Those were calculated following, \textit{e.g.}, Ref.~\cite{snyder_complex_2008}:
\begin{align*}
\eta = (1-\frac{\Tout}{\Tin})\frac{\sqrt{1+ZT}-1}{\sqrt{1+ZT}+\Tout/\Tin}.
\end{align*}

\end{document}